\documentclass[10pt,letterpaper]{article}
\usepackage[top=0.85in,left=2.75in,footskip=0.75in]{geometry}
\usepackage{comment}
\usepackage{amsmath,amssymb}
\usepackage{booktabs} 
\usepackage{changepage}

\usepackage{textcomp,marvosym}

\usepackage{cite}

\usepackage{nameref,hyperref}

\usepackage[right]{lineno}

\usepackage[nopatch=eqnum]{microtype}
\DisableLigatures[f]{encoding = *, family = * }

\usepackage[table]{xcolor}

\usepackage{array}

\usepackage{xurl}
\usepackage{hyperref}

\newcolumntype{+}{!{\vrule width 2pt}}

\newlength\savedwidth

\usepackage{setspace} 
\usepackage[aboveskip=1pt,labelfont=bf,labelsep=period,justification=raggedright,singlelinecheck=off]{caption}

\makeatletter
\renewcommand{\@biblabel}[1]{\quad#1.}
\makeatother

\usepackage{lastpage,fancyhdr,graphicx}
\usepackage{epstopdf}
\fancyheadoffset[L]{2.25in}
\fancyfootoffset[L]{2.25in}
\begin{document}
\vspace*{0.2in}


{\Large
\textbf\newline{Unveiling the Listener Structure Underlying K-pop’s Global Success: A Large-Scale Listening Data Analysis} 
}
\newline
\\
Ryota Nakamura \textsuperscript{1\dag, \ddag},
Keita Nishimoto \textsuperscript{1\dag*},
Ichiro Sakata \textsuperscript{1}
Kimitaka Asatani \textsuperscript{1},
\\
\\
\bigskip
\textbf{1} Graduate School of Engineering, The University of Tokyo, Japan
\bigskip

* Corresponding author: keita-nishimoto@g.ecc.u-tokyo.ac.jp\\
\dag Co-first author
\ddag The author is now with Preferred Networks, Inc., Japan.

\section*{Abstract}
From the mid-2000s to the 2010s, K-pop moved beyond its status as a regionally popular genre in Asia and established itself as a global music genre with enthusiastic fans around the world. However, little is known about how the vast number of music listeners across the globe have listened to and perceived K-pop. This study addresses this question by analyzing a large-scale listening dataset from Last.fm. An analysis of the distribution of play counts reveals that K-pop experienced a significant increase in plays between 2005 and 2019, largely supported by a small group of heavy listeners. The Gini coefficient in play counts is notably greater than that of existing mainstream genres and other growing niche genres. Furthermore, an analysis based on user-assigned genre tags quantitatively demonstrates that between 2005 and 2010, K-pop shed its status as a local Asian genre and established itself as a distinct music genre in its own right.

\clearpage
\section*{Introduction}
In recent years, K-pop has achieved remarkable success in the global music market, expanding its international influence. Its origins can be traced back to the late 1990s Korean Wave (Hallyu) boom, during which Korea’s cultural exports rapidly grew, especially in Asian markets. Since the 2000s, K-pop has expanded beyond Asia, and in 2012, PSY’s “Gangnam Style” became a global sensation, now boasting over 5 billion views on YouTube. This event marked a turning point in the worldwide recognition of Korean music \cite{sung2014k, madrid2015transatlantic}. In 2017, BTS won the “Top Social Artist” award at the Billboard Music Awards \cite{herman2017bts, uritoofab2017, cirisano2018bts, gibson2018k}.

The global spread of K-pop has been attributed to a combination of factors: the active participation of fandoms via new media such as social networking services (SNS), support from the Korean government \cite{hong2014birth, kim2016cultural, jin2016social, kwon2014cultural, ryoo2020cultural}, and strategic initiatives by Korean entertainment agencies \cite{shin2013organizing, cho2023leading}. In particular, K-pop fandoms function not merely as consumers but as active agents of international diffusion—sharing information on social media, producing viral content, and engaging in interactive communication with idols \cite{jang2017influences, cruz2021cultural, kang2022behind, jung2014social}. Furthermore, K-pop is characterized by its cultural hybridity, blending Western musical elements with traditional Korean styles and incorporating multilingual lyrics, thereby establishing itself as a unique musical genre with global competitiveness \cite{shim2006hybridity, oh2009hallyu, jin2014critical, yoon2018global}.

In recent years, K-pop acts have frequently appeared at the top of music charts in various countries, and fandom activities via SNS have been noted for their high level of engagement \cite{kim2014comparative, zhang2020east}. On the other hand, some have pointed out that K-pop’s popularity is heavily reliant on a small number of highly enthusiastic fans \cite{liu2025visibility}, and that its reach may not extend widely to the general listening public in some countries \cite{min2019transcultural, kanozia2021cultural, yoon2022between}. However, existing studies rely mainly on qualitative approaches such as interviews and questionnaires and therefore have not provided clear answers to these debates, highlighting the need for quantitative research that measures the extent to which K-pop is actually accepted as music across different countries.

In summary, previous research has primarily explained the popularity of K-pop through four lenses: (1) fandom activity centered on social media \cite{jang2017influences, cruz2021cultural, kang2022behind, jung2014social}, (2) government support for the cultural industry \cite{hong2014birth, kim2016cultural, jin2016social, kwon2014cultural, ryoo2020cultural}, (3) talent development systems and global expansion strategies of entertainment agencies \cite{shin2013organizing, cho2023leading}, and (4) the cultural hybridity of K-pop \cite{shim2006hybridity, oh2009hallyu, jin2014critical, yoon2018global}. In contrast, this study focuses on how music listeners engage with K-pop, analyzing both its diffusion among general listeners and its perceived position within the broader music landscape.

Specifically, this study addresses the following research questions concerning how K-pop established its global popularity, using the listening histories of approximately 50,000 users from the online music service Last.fm dataset \cite{schedl2022lfm} spanning from 2005 to 2019.

\begin{enumerate}
  \item \textbf{RQ1: Has K-pop been received to a similar extent as existing music genres?}
  
  Although K-pop has gained international competitiveness—frequently appearing on charts such as the Billboard Hot 100—there are very few academic studies that quantitatively demonstrate the extent to which it has spread and been accepted among music listeners. Some existing studies suggest that its popularity is supported primarily by specific regions or fan groups \cite{liu_visibility_2025, lee_how_2020}. Has K-pop been listened to as widely as mainstream genres like pop, rock, or rap, or is it predominantly consumed by a small subset of listeners? To clarify this, we compare K-pop with other music genres that were originally considered niche or local but have recently gained global traction—such as Chillwave and Djent.

  \item \textbf{RQ2: What types of musical preferences characterize K-pop listeners, and how is K-pop perceived as a genre?}
  In addition to RQ1, this study investigates the musical preferences of listeners who actively enjoy K-pop. While prior studies have explored aspects such as the racial and sexual identities of K-pop fans \cite{kuo2022performance, shin2018queer}, research on their musical tastes remains limited. We examine whether K-pop is favored by listeners with a narrow set of musical interests, or whether it appeals to a wide range of listeners with diverse genre preferences. Moreover, Last.fm allows users to freely assign genre tags to tracks, enabling an analysis of how K-pop is perceived in genre terms. Is K-pop understood as part of the broader category of Asian music genres, like J-pop (Japanese pop), or has it acquired an independent identity as its own genre? We explore this by comparing K-pop with J-pop, another prominent genre of Asian pop music.

\end{enumerate}

Clarifying these two research questions not only helps retrace the trajectory of K-pop's global ascent, but also provides insight into the broader process by which a local and niche music genre can achieve worldwide diffusion, yielding important implications for the future of the music industry.

\section*{Methods and Data}
\subsection*{Last.fm Dataset and Pre-processing}
This study utilizes the LFM-2b dataset \cite{schedl2022lfm}, which includes listening histories from the online music service Last.fm. Constructed as a successor to LFM-1b \cite{schedl2016lfm}, the LFM-2b dataset contains 2,014,164,872 listening records generated by over 120,000 users between February 14, 2005, and March 20, 2020. It also includes users’ demographic information—such as country, age, and gender—as well as tag information attached to tracks.

Last.fm aggregates listening logs from various platforms including Spotify and YouTube, allows users to share their listening activity, and incorporates a music recommendation system. For this study, we use three components of the dataset: (1) listening-events, (2) users, and (3) tags-micro-genres—a subset of 2,808 genre-related tags extracted from a total of 1,041,819 user-generated tags, excluding non-musical descriptors.

The (1) listening-events data contains records of user listening behavior in the format <user ID, artist ID, track ID, album ID, timestamp>. The timestamp indicates the listening start time in seconds. To protect user privacy, only numeric user IDs are retained in the public version of the dataset. Tracks included in the LFM-2b dataset are limited to those that have been played more than 10 times.

The (2) users data includes demographic details for 120,332 individuals, such as country, age, gender, and account creation time. Country information is represented using ISO 3166 Alpha-2 codes. A total of 65,136 users have no recorded country information. The account creation time corresponds to the moment a user created their profile.

The (3) tags-micro-genres data represents genre-related tags that users have attached to tracks. Based on the classification system of the genre taxonomy project Every Noise at Once \cite{everynoise}, only tags recognized as music genres are retained. Each track is annotated with a dictionary of (tag, weight) pairs. The weight is an integer from 1 to 100, rounded to the nearest whole number. The tag with the highest frequency on a given track is assigned a weight of 100, and all other tags receive weights proportionally scaled to their relative frequency compared to the most common tag. In analyses that utilize tag weights in this study, the weights assigned to each track are normalized such that their sum equals one. 
The tags-micro-genres data contains 2,807 unique tags, with a total of 4,978,282 tag assignments. In addition to major genres such as rock, pop, jazz, folk, and hip hop, niche and atmospheric genres like metal, ambient, and soundtrack are also highly represented, indicating that Last.fm users assign tags across a broad spectrum of musical styles (S1 Fig (c)).

For this study, we first excluded tracks without any tags. Then, for each track, the tag with the highest weight was defined as the \textit{representative tag} and treated as the genre classification of that track. As a result, we obtained 1,019,352,751 listening events covering the period from February 14, 2005, to December 31, 2019. Listening records from 2020 were excluded due to incomplete coverage—only available up to March 20. The dataset includes 55,177 unique users, with only 18 users lacking country information. The number of users peaked in 2012 and declined afterward (S1 Fig (a)). The user base skews male and is concentrated in North America, South America, and Europe, with relatively limited participation from Asia and Africa (S1 Fig (b)).

As the release year of each track is not included in the Last.fm dataset, we approximated it by using the first year in which each track was listened to. A random sampling check (n = 30) revealed that in approximately 80\% of cases (n = 25), this proxy matched the actual release year. About 10\% (n = 4) differed by one year, and one outlier showed a 13-year discrepancy. Nevertheless, this method is considered sufficiently practical for the purposes of this study.

\subsection*{Emerging Niche Genres}
As a basis for comparison with K-pop, we define a set of \textit{emerging niche genres} by extracting the top five music genres (K-pop, Chillwave, Djent, Trap, New Rave) with the highest cumulative play count growth rates from 2005 to 2019 (see S2 Fig). The growth rate for each genre was calculated by subtracting the cumulative number of plays in the genre’s ``birth year" from the cumulative number of plays in 2019, then dividing by the number of years elapsed. In this context, the birth year of a genre is defined as the first year in which more than 10 users listened to that genre.

\subsection*{Musical Preference Vectors}
To capture individual listeners’ musical preferences, we aggregated each user’s play counts by genre and transformed these counts into vectors using term-frequency–inverse-document-frequency (TF-IDF) weighting; we refer to the resulting representations as \textit{musical preference vectors}. Applying K-means clustering to these vectors allows us to partition listeners into groups that exhibit similar patterns of musical preference.


\section*{Results}
\subsection*{Growth Trajectory of K-pop}
\begin{figure}[!h]
\centering
\includegraphics[width=1.0 \linewidth]{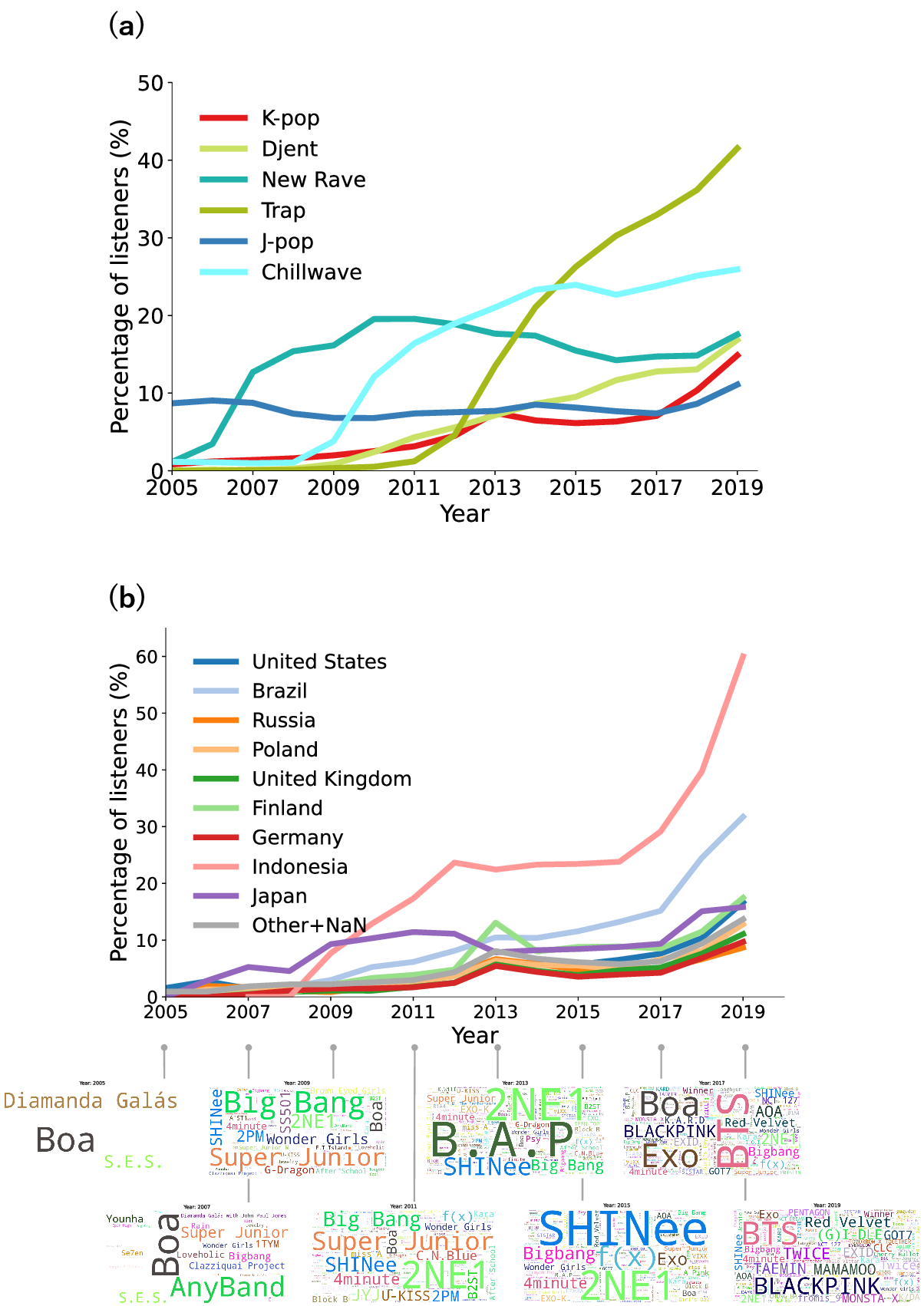}
\caption{\textbf{Trends in adoption of niche genres and K-pop among listeners.} \\
(a) Yearly trends in the proportion of all listeners who played at least one track from any of the emerging niche genres.
(b) Yearly trends in the proportion of listeners in each country who played at least one K-pop track. Additionally, word clouds show the most frequently played K-pop artists for each year based on play counts.}
\label{fig1}
\end{figure}

Fig \ref{fig1}(a) shows the yearly trends in the proportion of listeners who played at least one track belonging to K-pop or other emerging niche genres. Fig \ref{fig1}(b) illustrates the trends in the proportion of listeners who played K-pop at least once in each selected country, along with a WordCloud representing the most-played K-pop artists in each year. For the country selection, we included the top seven countries by user count (from Europe and the Americas), as well as the top two countries from the Asia region (Japan and Indonesia). Note that South Korea is excluded from the country-level analysis due to the small number of Last.fm listeners located there. Furthermore, Figure \ref{fig1}(b) reveals that in 2019 approximately 60\% of Indonesian users streamed K-pop—by far the largest share among the countries examined. This estimate should be interpreted with caution, however, because Indonesia also has the smallest registered listener base in the sample, a factor that can inflate the observed proportion.

As of 2023, K-pop is typically categorized into four generations: the first generation (late 1990s to early 2000s), the second generation (around 2001 to around 2011), the third generation (around 2012 to around 2018), and the fourth generation (2019 onward) \cite{ryou_generations_2023}.

In 2005, first-generation artists such as S.E.S and BoA were being played (The artist “Diamanda Galás,” who is not originally associated with K-pop, appears in some of the WordClouds because users assigned the “K-pop” tag to certain her tracks).
Subsequently, from around 2007 to 2013, the popularity shifted to second-generation groups such as SUPER JUNIOR, Wonder Girls, 2NE1, SHINee, BIGBANG, and B.A.P, along with an increase in the number of listeners. Although Psy’s “Gangnam Style” became a worldwide hit in 2012, it was not heavily played according to Last.fm data, resulting in Psy appearing relatively small in the WordCloud. During this period, K-pop first gained listeners in Japan, followed by Indonesia, then Brazil, Europe, and the United States. It has been suggested by the existing study \cite{cho_k-pop_2023} that K-pop first became popular in Asia, including Japan and Indonesia, during the 2000s, and then expanded to Europe and the North/South Americas in the 2010s and beyond. The results obtained here are largely consistent with that progression.

After 2013, while second-generation artists continued to have a strong presence, third-generation groups such as EXO, BTS, and Red Velvet gradually increased their prominence in the WordCloud. The growth in the number of Brazilian listeners that began around 2009 continued beyond 2013.

Between 2013 and 2016, the overall proportion of K-pop listeners remained relatively stagnant. However, around 2017, a notable increase was observed, alongside a surge in play counts for third-generation groups such as BTS and BLACKPINK. Particularly, BTS has been reported as a major catalyst for the global expansion of K-pop’s popularity \cite{parc_analyzing_2020}, and this influence is also reflected in the current results.

These findings suggest that the Last.fm data broadly captures the real-world trends in the popularity of K-pop and its associated artists.

\subsection*{Has K-pop been received to a similar extent as existing music genres?}
\begin{figure}[!h]
\centering
\includegraphics[width=0.7 \linewidth]{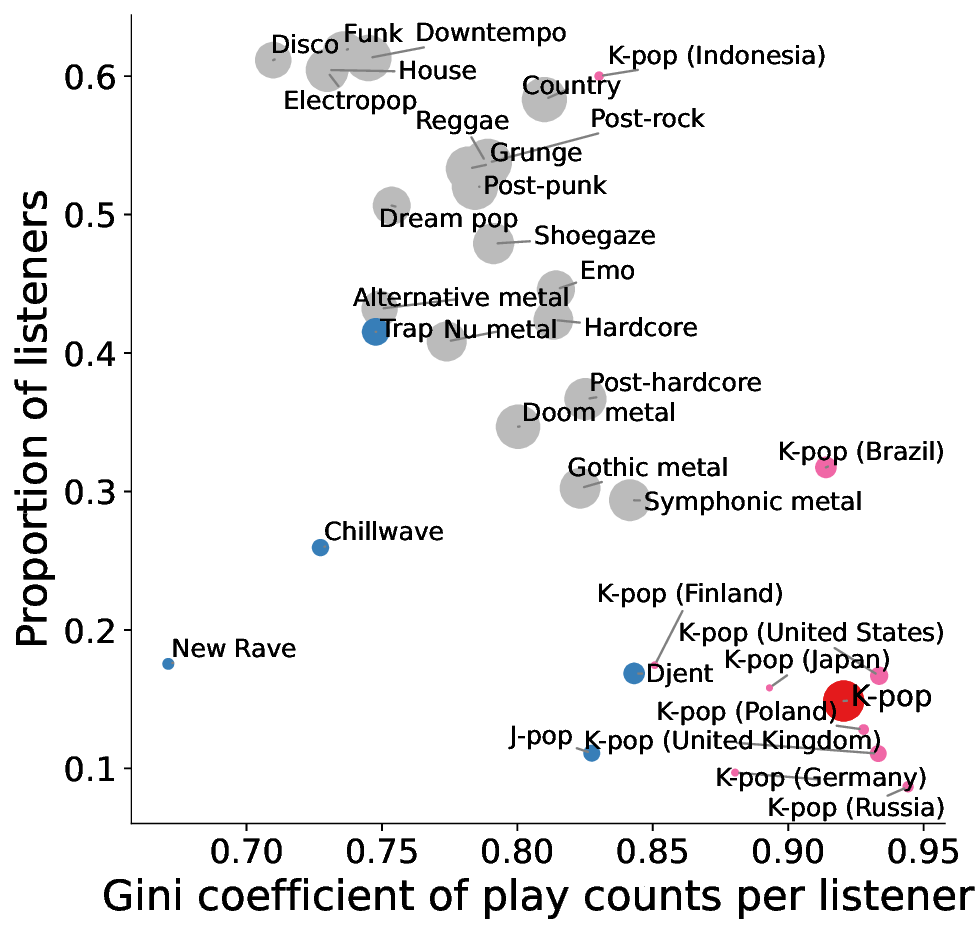}
\caption{\textbf{Listener proportion and concentration of play counts across music genres and countries.} \\
The vertical axis represents the proportion of users who listened to each genre at least once in 2019 (listener proportion), while the horizontal axis shows the Gini coefficient of play counts per user for that genre. The size of each circle corresponds to the total number of plays. K-pop is shown in red and pink, existing mainstream genres in gray, and emerging niche genres in blue.}
\label{fig2}
\end{figure}

Next, to address RQ1 — Has K-pop been received to a similar extent as existing music genres? — Fig \ref{fig2} presents a comparison between K-pop, existing mainstream genres, and other emerging niche genres in terms of their listener proportion (i.e., the percentage of users who listened to at least one track from a given genre in 2019; vertical axis) and the Gini coefficient of play counts per listener (horizontal axis).
The size of each circle represents the total number of plays for the corresponding genre.
K-pop is represented by a red circle, other emerging niche genres by blue circles, and existing mainstream genres by gray circles.
The existing mainstream genres were selected by identifying the ten genres ranked immediately above and below K-pop in terms of total play counts, thereby capturing genres of comparable popularity.
In addition, to analyze K-pop’s reception across countries, we include pink-colored circles labeled “K-pop (Country Name),” which indicate the listener proportion and Gini coefficient of K-pop within individual countries.

The figure clearly shows that, compared to existing genres with similar levels of popularity and play counts, K-pop has a lower proportion of listeners but a higher Gini coefficient. This suggests a strong skew in the distribution of play counts among users. Even compared to other emerging niche genres (blue circles), K-pop’s Gini coefficient stands out, further highlighting its uniqueness. Additionally, when examining individual countries, K-pop generally maintains a high Gini coefficient across nations, with the exception of Indonesia. Surprisingly, even in the United States—where BTS gained significant popularity in 2018—K-pop exhibits an extremely high Gini coefficient.

\begin{figure}[!h]
\centering
\includegraphics[width=0.7 \linewidth]{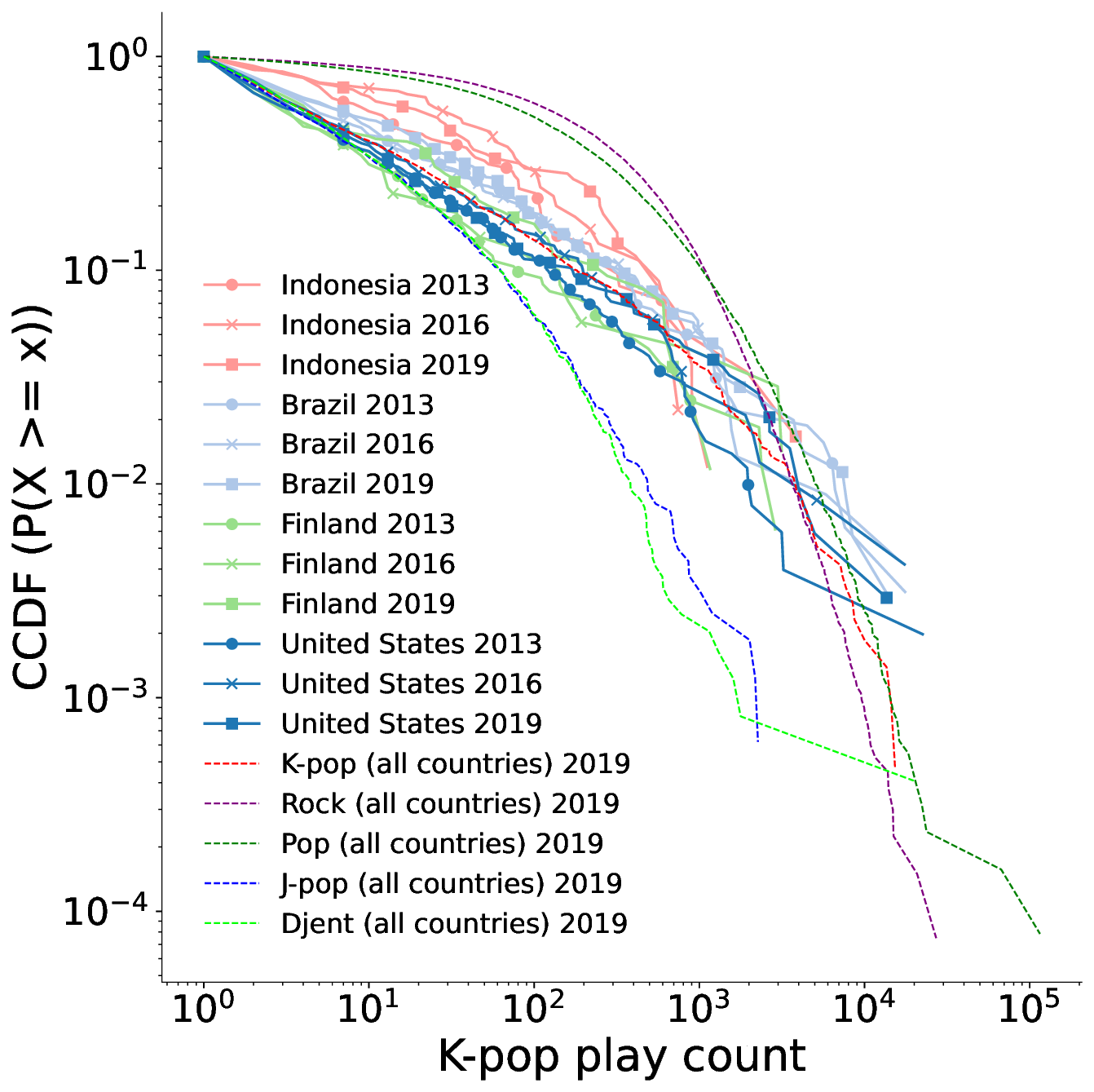}
\caption{\textbf{Cross-country and temporal distributions of individual-level K-pop play counts.}\\
Complementary Cumulative Distribution Functions (CCDFs) of listener-level K-pop play counts for each year in Indonesia, Brazil, Finland, and the United States.}
\label{fig3}
\end{figure}

Fig \ref{fig3} further explores this pattern by presenting the complementary cumulative distribution functions (CCDFs) of listener play counts in 2013, 2016, and 2019 for the four countries with the highest domestic K-pop listener ratios in 2019 (see Fig \ref{fig1}(b)). As a point of comparison, CCDFs for other popular genres (Pop, Rock) and emerging genres (Djent, J-pop) in 2019 are also shown.

\begin{figure}[!h]
\centering
\includegraphics[width=0.65 \linewidth]{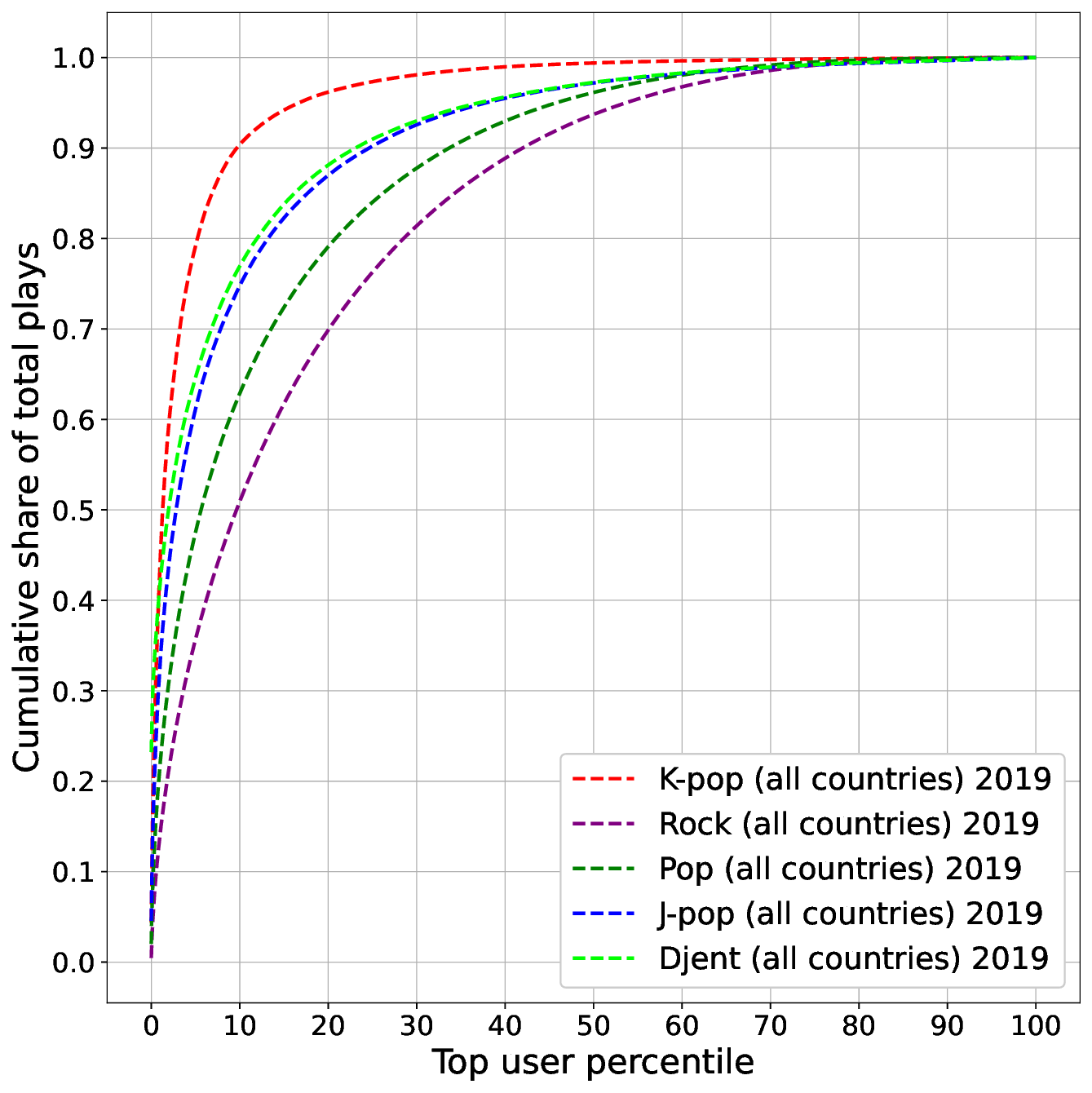}
\caption{\textbf{Concentration of music listening activity among users by genre.}\\
This figure shows the concentration of music playback activity among users for each music genre. The horizontal axis represents the top x\% of users, ranked by individual total play count, while the vertical axis indicates the cumulative share of total plays accounted for by these users.}
\label{fig4}
\end{figure}

Except for Indonesia, the distribution of K-pop play counts across countries follows a heavy-tailed distribution, resembling either a log-normal or power-law distribution (see S3 Table and S4 Fig), clearly differing from the lighter-tailed distributions observed for other music genres. Although the slope of the CCDF, and thus the shape of the distribution, varies somewhat by country, it remains relatively stable across years, suggesting that the skewed distribution of play counts is a robust feature. This is further supported by Fig \ref{fig4}, which plots the relationship between user percentiles and cumulative play counts: for K-pop, play counts are highly concentrated among top users compared to other genres. Specifically, the top 10\% of users account for approximately 90\% of the total play counts.

Taken together, these results indicate that, across many countries, K-pop has exhibited a highly skewed distribution of play counts compared to other genres, with its popularity strongly supported by a relatively small group of heavy listeners.

\subsection*{What types of musical preferences characterize K-pop listeners, and how is K-pop perceived as a genre?}
\begin{figure}[!h]
\centering
\includegraphics[width=1.0 \linewidth]{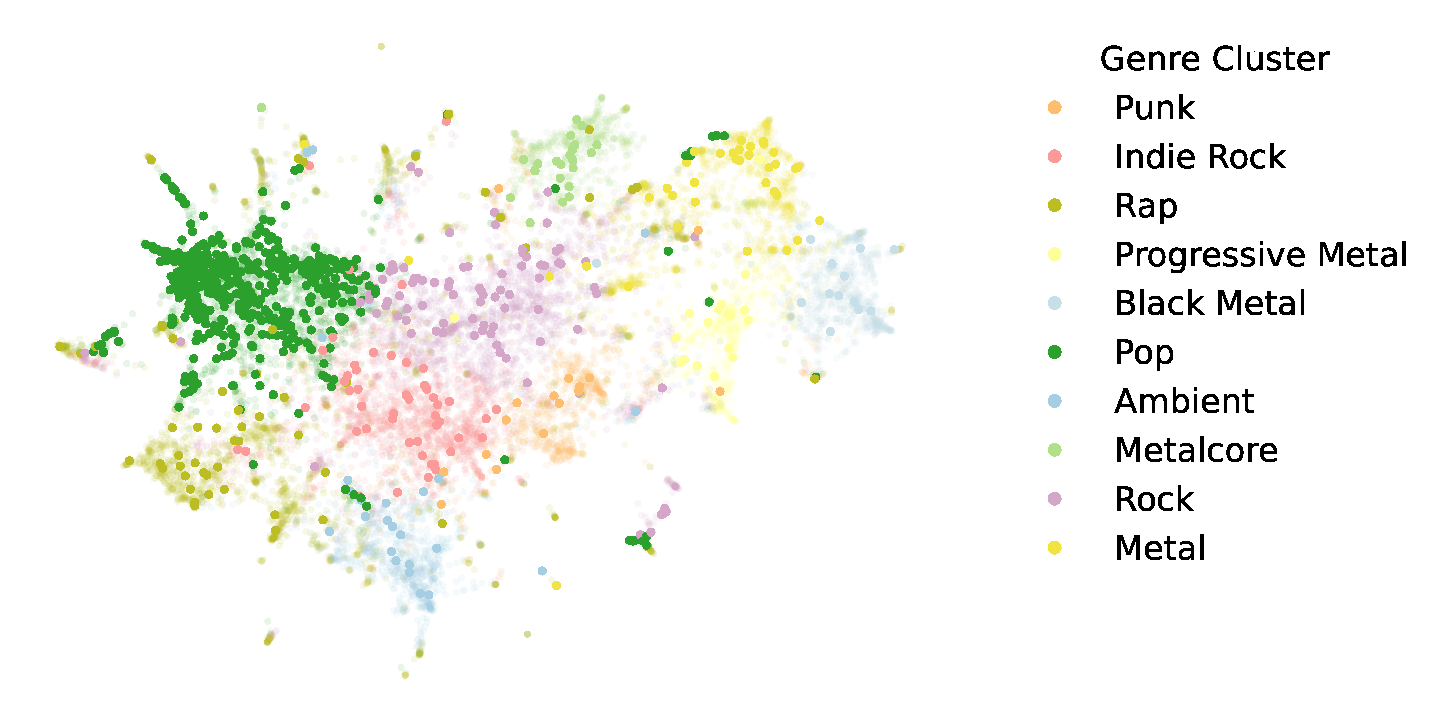}
\caption{\textbf{Clustering of users based on genre preferences and K-pop heavy listeners.}\\
The plot shows a two-dimensional UMAP projection of k-means clustering (k = 10) results based on listeners' genre preference vectors from 2019.
The darker-colored points represent heavy listeners of K-pop — defined as those who listened to K-pop more than 10 times in 2019.}
\label{fig5}
\end{figure}

In RQ1, it was revealed that K-pop play counts are highly concentrated among a small group of listeners. We now turn to RQ2: What types of musical preferences characterize K-pop listeners, and how is K-pop perceived as a genre?
Fig \ref{fig5} presents the results of clustering users based on their musical preference vectors using k-means clustering (with the number of clusters set to 10), followed by mapping the results into a two-dimensional space using UMAP \cite{mcinnes_uniform_2020}. Users shown in dark colors represent heavy K-pop listeners, defined as those who played K-pop tracks more than 10 times within a year. Cluster names are assigned based on the music genre that dominates the listening histories of users belonging to each cluster.

From the figure, it is evident that heavy K-pop listeners are concentrated in the Pop listener cluster. Specifically, 42\% of heavy K-pop listeners belong to the Pop cluster (the largest proportion among all clusters), and 10\% of users within the Pop cluster are heavy K-pop listeners (also the highest among all clusters).

\begin{figure}[!h]
\centering
\includegraphics[width=0.9 \linewidth]{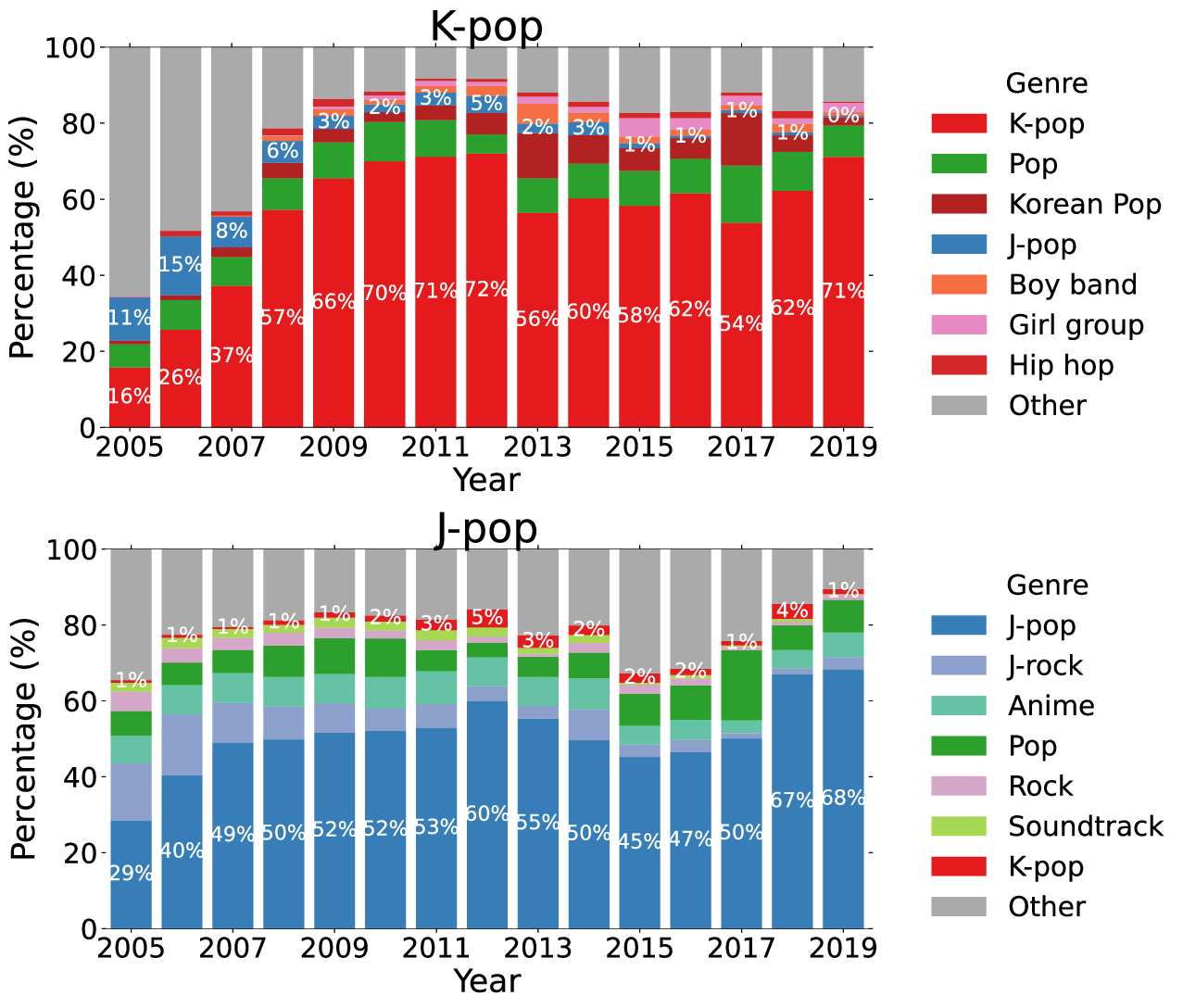}
\caption{\textbf{Temporal changes in associated tag usage for K-pop and J-pop tags.}\\
Proportion of tags assigned to tracks containing K-pop/J-pop tags, aggregated by release year. The proportion of tag $t$ is computed as the total normalized weight of tag $t$ across the tracks $\mathcal{T}$ containing the target tag, divided by the total tag weight across all tags $\mathcal{A}$ and target tracks: ${\sum_{i \in \mathcal{T}} w_{t,i}}/{\sum_{s \in \mathcal{A}} \sum_{i \in \mathcal{T}} w_{s,i}}$, where $w_{t,i}$ is the normalized weight of tag $t$ contained in a track $i$.}, 
\label{fig6}

\includegraphics[width=0.65 \linewidth]{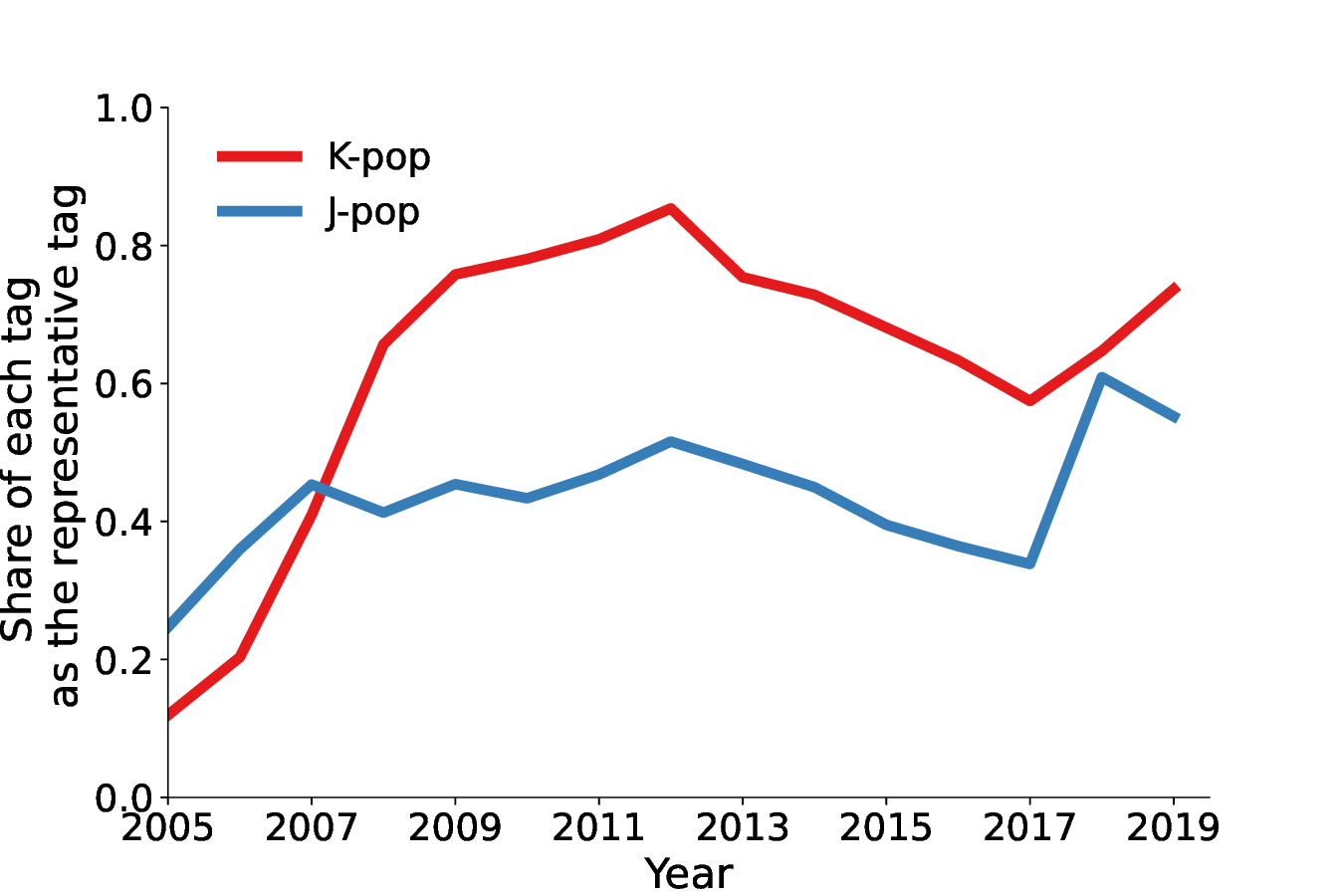}
\caption{\textbf{Temporal trends in the use of K-pop and J-pop as representative tags.}\\
Trends in the proportion of tracks for which each tag serves as the representative tag among tracks containing the K-pop or J-pop tag. 
The horizontal axis represents the release year of the tracks.}
\label{fig7}
\end{figure}

To address RQ2—how is K-pop perceived as a genre?—we analyzed the changes in tags assigned to K-pop tracks and conducted a comparative analysis with J-pop (Japanese pop).
It should be noted that, in order to capture situations where the recognition of K-pop/J-pop was still low, we included tracks where K-pop/J-pop was not necessarily the representative tag.
We extracted tracks that contained either K-pop or J-pop tags and, considering the user-assigned tag weights, examined the yearly trends in the proportion of tags assigned to tracks that debuted each year (Fig \ref{fig6}). 
Furthermore, Fig \ref{fig7} shows the trends in the proportion of tracks for which each tag (K-pop/J-pop) was the representative tag.

From Fig \ref{fig6}, it is evident that among tracks containing the K-pop tag, the proportion of tracks labeled with K-pop as the representative tag surged rapidly between 2005 and 2010. Although a similar upward trend (from 29\% to 52\%) was observed for J-pop, the increase for K-pop (from 16\% to 70\%) was notably steeper. 
Simultaneously, the proportion of J-pop tags among tracks containing the K-pop tag gradually declined.
Moreover, among tracks that included the K-pop tag, the proportion with K-pop as the representative tag increased to approximately 80\%, whereas the corresponding proportion for J-pop only rose to about 50\%.

These results suggest that K-pop, originally perceived as a miscellaneous genre similar to J-pop, established a distinct and independent identity from 2005 to 2010. The difference in the proportion of representative tags indicates that while J-pop was often attached as a secondary tag alongside others, K-pop was more frequently used as the primary representative tag for tracks, implying a higher degree of genre independence for K-pop.

An interesting observation is that between 2013 and 2017, both the proportion of tracks tagged with K-pop (see Fig \ref{fig6}) and the proportion of tracks with K-pop as the representative tag declined (see Fig \ref{fig7}), before rebounding to previous levels in 2018–2019. This trend aligns with the period of decline and subsequent resurgence in K-pop listener proportions shown in Fig \ref{fig1}(a). It suggests that during this period, K-pop experienced a temporary stagnation in popularity and a decline in its presence as a distinct genre, followed by a resurgence driven by the success of third-generation artists such as BTS and BLACKPINK.

\section*{Discussion}
This study investigated two research questions regarding the acceptance and diffusion of K-pop:
RQ1: Has K-pop been received to a similar extent as existing music genres?
RQ2: What types of musical preferences characterize K-pop listeners, and how has K-pop been perceived as a genre?
To address these questions, we conducted an analysis using the large-scale listening dataset from online music service Last.fm.

Regarding RQ1, our analysis revealed that during the period up to 2019, K-pop, with the exception of Indonesia, had not become widespread among the general listener population in many countries. Instead, it was strongly supported by a relatively small group of heavy listeners. This tendency was consistently observed across different years and countries (again, excluding Indonesia). Furthermore, the level of skewness in play counts, as measured by the Gini coefficient, was higher for K-pop not only compared to existing music genres with similar play counts, but also compared to other representative emerging niche genres.
The unique pattern observed in Indonesia is likely due to two factors: Indonesia had the smallest number of users among the selected countries, and K-pop has achieved a particularly broad level of acceptance there, as evidenced by BTS having the highest monthly YouTube view counts and Instagram follower numbers in Indonesia compared to any other country \cite{liu2025visibility}. 

Regarding RQ2, the results showed that K-pop was mainly embraced by listeners who also favor Pop music. The analysis of user-assigned tags further suggested that between 2005 and 2010, K-pop transitioned from being perceived as just another Asian music genre to establishing itself as an independent genre. This period coincides with the time when first-generation artists such as S.E.S and BoA, along with second-generation groups such as SUPER JUNIOR, BIGBANG, and Wonder Girls, were frequently listened to. These findings imply that these artists played a significant role in the formation of K-pop’s distinct identity.

Why, then, is the skewness in K-pop play counts so pronounced? One plausible explanation is the effective cultivation of heavy users—now often referred to as superfans \cite{noauthor_superfans_2023}—within the K-pop ecosystem. A defining feature of K-pop fandom is its participatory approach \cite{ismail_kpop_2023}, where fans not only passively listen to music but also engage in various active support activities such as posting translation videos, cover performances, and remixes on YouTube, as well as participating in donation campaigns addressing social issues \cite{lee_how_2020}.
Previous studies have also reported that organized mass streaming aimed at boosting chart rankings is a distinctive feature of K-pop fandom behavior \cite{lee_how_2020, chung_bts_2022}. The skewed play count distribution observed in this study likely reflects such activities.
In recent years, the importance of superfans has been increasingly recognized within the music industry \cite{noauthor_superfans_2023}, and the findings of this study support this emerging trend.

Since 2005, the proportion of tracks that list K-pop among their tags and adopt K-pop as their representative tag has climbed sharply, settling at roughly 60–80 \% after 2010—consistently higher than the corresponding share for J-pop over the same period. These results suggest that, while K-pop songs appeal to listeners with a preference for pop music and stylistically exhibit considerable hybridity, audiences do not treat K-pop as a sub-branch of another genre; rather, they recognize and consume it as a distinct, autonomous genre in its own right.

Despite these contributions, this study has several limitations. The most significant is the limitation inherent in the Last.fm dataset. The listening data used in this study only extends up to December 2019. To mitigate this limitation, especially in addressing RQ1, we compared K-pop with both existing mainstream genres with similar play counts and emerging niche genres, and conducted analyses by observation year and country, obtaining consistent results.
Nonetheless, it remains possible that trends have shifted in the current period, especially considering the impact of the COVID-19 pandemic that began in 2020 and the emergence of fifth-generation K-pop artists.

Another point of caution concerns the listener demographics. It has been reported that K-pop fans are predominantly young women who are active on social media platforms \cite{laffan_positive_2021}. This demographic differs from the user base of Last.fm, where male users make up approximately 70\% of the population, and the platform itself targets music enthusiasts. Moreover, the appeal of K-pop extends beyond music to visual and performative aspects, including music videos, concerts, and fan interactions \cite{ismail_kpop_2023}.
This study focused specifically on listening behavior, which is only one facet of K-pop fandom. To fully capture the comprehensive entertainment experience that K-pop offers, future research should incorporate other dimensions beyond listening data.

Despite these limitations, this study quantitatively demonstrated, for the first time using large-scale listening data, that between 2005 and 2019, (i) K-pop’s global rise was strongly supported by participatory fandom activities and superfans, and that (ii) K-pop achieved a high degree of independence as a musical genre.
While it remains uncertain how essential these two elements are to the global expansion of other local and niche music genres, the insights gained here provide a valuable contribution toward understanding the mechanisms by which local genres can grow into global phenomena.

\newpage
\bibliography{reference}

\newpage
\nolinenumbers
\section*{Supporting information}

\paragraph*{S1 Fig.}
Basic analyses of our pre-processed dataset: (a) Trend in the number of unique users, (b) Top 25 countries by registered users, and (c) Genre distribution of tracks contained.
\begin{figure}[!h]
\centering
\includegraphics[width=1.0\linewidth]{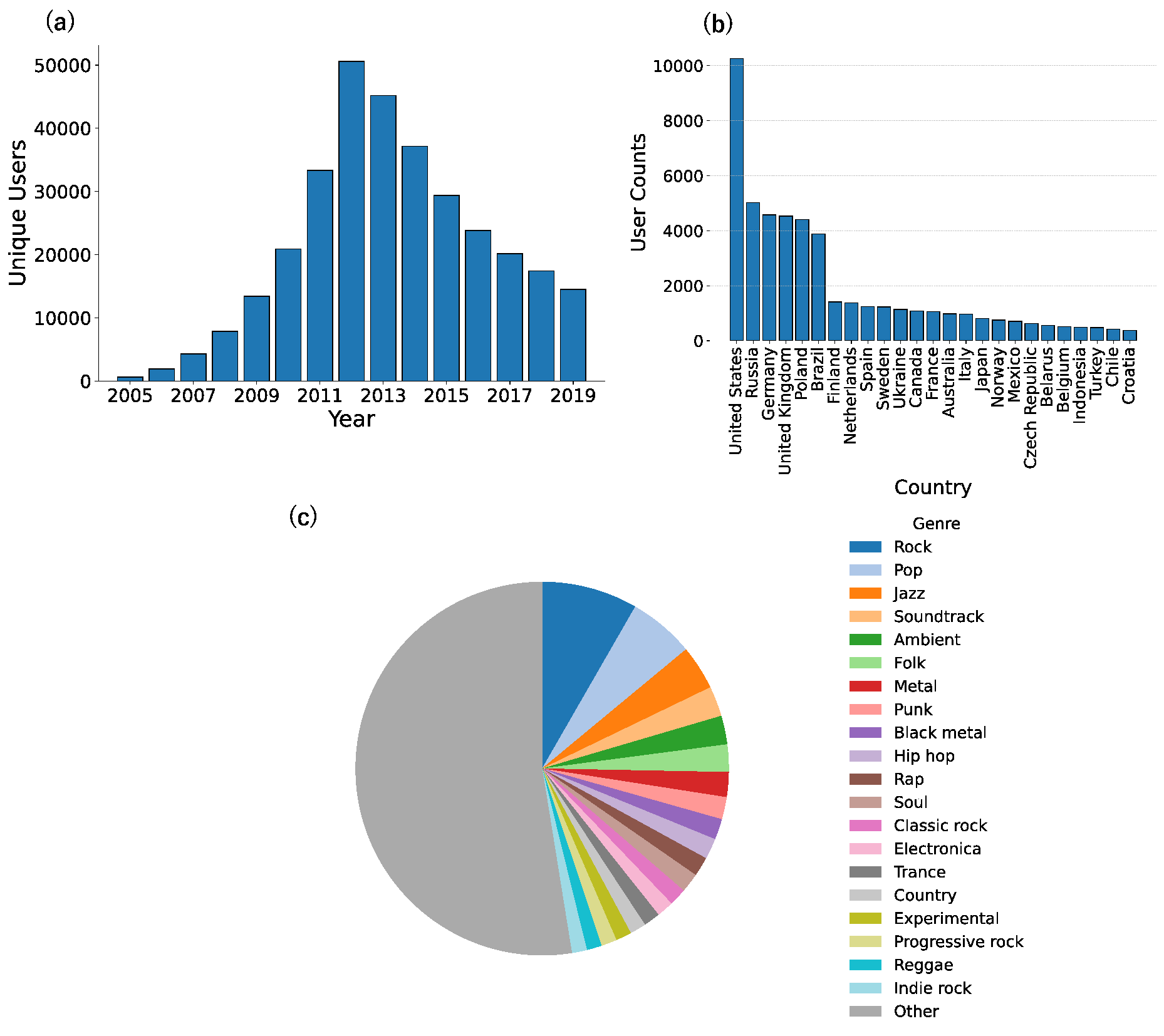}
\end{figure}

\newpage
\paragraph*{S2 Fig.}
Trends in cumulative play counts of \textit{emerging niche genres}. Emerging niche genres are selected as the top five music genres with the highest growth rates in cumulative play counts.
\begin{figure}[!h]
\centering
\includegraphics[width=0.8 \linewidth]{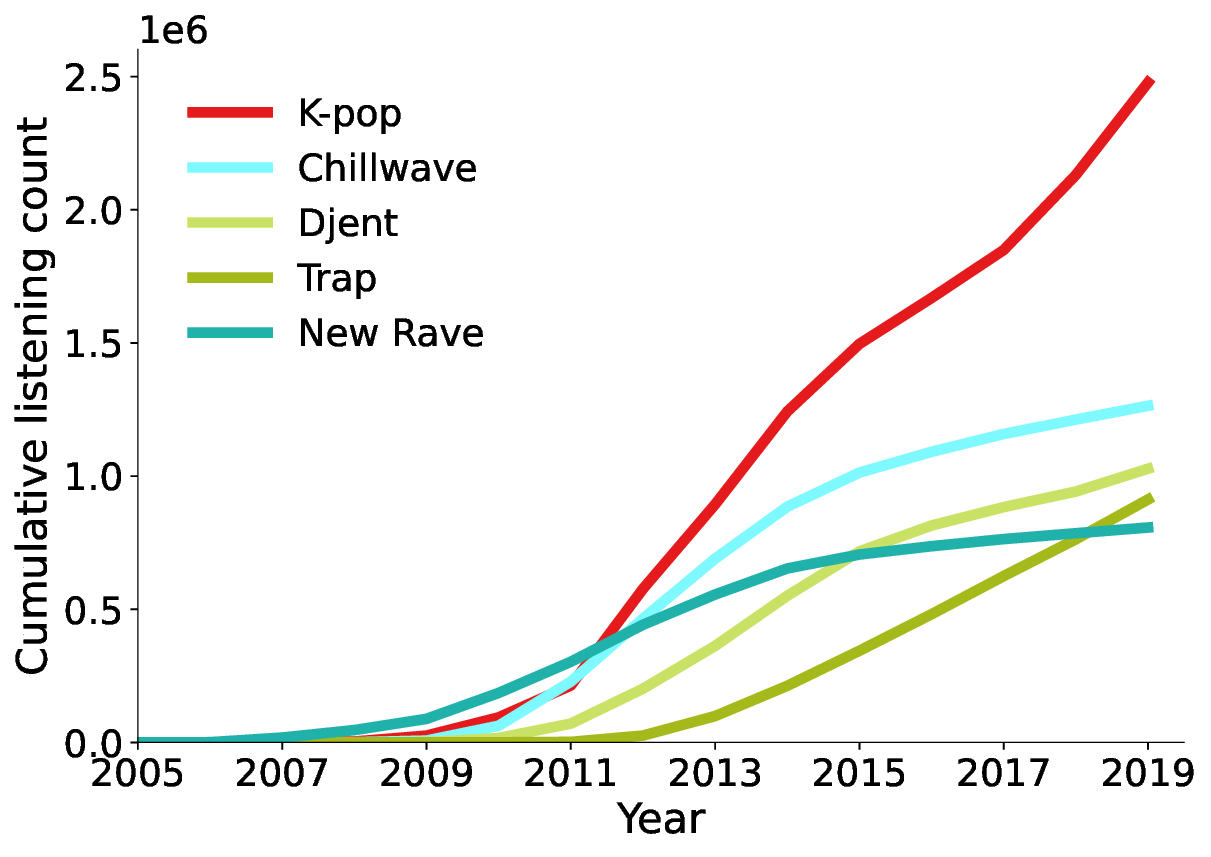}
\label{fig_s2}
\end{figure}

\newpage
\paragraph*{S3 Table.}
The table shows the results of likelihood ratio tests comparing the goodness-of-fit of three candidate distributions—power law (PL), exponential (Exp), and lognormal (Logn)—to listener play count data from four countries over three time points (2013, 2016, and 2019). 
We used powerlaw library \cite{alstott2014powerlaw} for calculation.
Each row reports the log-likelihood ratio ($R$), corresponding $p$-value, $z$-value and sample size ($n$) for a pairwise comparison between two models, as indicated in the “vs” column. 
The interpretation is based on the sign of $R$ and whether $p < 0.017$ (adjusted for multiple comparisons using the Bonferroni correction): a significantly positive $R$ indicates that the first model listed in the comparison provides a better fit, while a significantly negative $R$ favors the second model. 
“The best-fitting model was selected based on pairwise likelihood ratio tests with Bonferroni-adjusted significance threshold ($p < 0.017$). If no significant differences were found among all three models, the entry is labeled ‘No significance.’ If two models were significantly better than a third but not distinguishable from each other, both are reported (e.g., ‘Power law or Lognormal’).”

\begin{table}
\centering
\label{tab:distribution_comparison_z_n_corrected}
\resizebox{\textwidth}{!}{
\begin{tabular}{cccccccl}
\toprule
Country & Year & vs & $R$ & $p$-value & $z$-value & $n$ & Fittest model \\
\midrule
Indonesia & 2013 & PL vs Exp & 3.4735 & 0.2852 & 1.0687 & 83 & No significance \\
          &      & PL vs Logn & -0.3541 & 0.5659 & -0.5742 & 83 & \\
          &      & Logn vs Exp & 3.8277 & 0.1473 & 1.4491 & 83 & \\
Indonesia & 2016 & PL vs Exp & 1.4781 & 0.5903 & 0.5383 & 45 & No significance \\
          &      & PL vs Logn & -0.5337 & 0.4937 & -0.6844 & 45 & \\
          &      & Logn vs Exp & 2.0118 & 0.3080 & 1.0195 & 45 & \\
Indonesia & 2019 & PL vs Exp & 4.8534 & 0.0692 & 1.8171 & 60 & No significance \\
          &      & PL vs Logn & 0.0001 & 0.9804 & 0.0246 & 60 & \\
          &      & Logn vs Exp & 4.8533 & 0.0688 & 1.8196 & 60 & \\
Brazil & 2013 & PL vs Exp & 51.5731 & 0.0010 & 3.3025 & 320 & Power law or Lognormal \\
       &      & PL vs Logn & -1.7055 & 0.2269 & -1.2085 & 320 & \\
       &      & Logn vs Exp & 53.2785 & 0.0002 & 3.6836 & 320 & \\
Brazil & 2016 & PL vs Exp & 414.3437 & 0.0000 & 5.5029 & 224 & Lognormal \\
       &      & PL vs Logn & -11.0140 & 0.0006 & -3.4221 & 224 & \\
       &      & Logn vs Exp & 425.3576 & 0.0000 & 5.8108 & 224 & \\
Brazil & 2019 & PL vs Exp & 82.2233 & 0.0000 & 4.4610 & 352 & Power law or Lognormal \\
       &      & PL vs Logn & -1.1786 & 0.2943 & -1.0488 & 352 & \\
       &      & Logn vs Exp & 83.4020 & 0.0000 & 4.7620 & 352 & \\
Finland & 2013 & PL vs Exp & 296.0363 & 0.0000 & 6.7054 & 163 & Power law or Lognormal \\
        &      & PL vs Logn & -3.2802 & 0.0391 & -2.0635 & 163 & \\
        &      & Logn vs Exp & 299.3166 & 0.0000 & 6.9995 & 163 & \\
Finland & 2016 & PL vs Exp & 147.2102 & 0.0000 & 5.2346 & 70 & Power law or Lognormal \\
        &      & PL vs Logn & -1.2442 & 0.2853 & -1.0685 & 70 & \\
        &      & Logn vs Exp & 148.4544 & 0.0000 & 5.4855 & 70 & \\
Finland & 2019 & PL vs Exp & 100.0320 & 0.0000 & 5.5857 & 85 & Power law or Lognormal \\
        &      & PL vs Logn & -4.3705 & 0.0189 & -2.3472 & 85 & \\
        &      & Logn vs Exp & 104.4026 & 0.0000 & 6.4455 & 85 & \\
United States & 2013 & PL vs Exp & 976.8546 & 0.0000 & 5.1982 & 505 & Lognormal \\
              &      & PL vs Logn & -15.6553 & 0.0000 & -4.1856 & 505 & \\
              &      & Logn vs Exp & 992.5098 & 0.0000 & 5.3473 & 505 & \\
United States & 2016 & PL vs Exp & 514.2970 & 0.0000 & 5.3333 & 238 & Lognormal \\
              &      & PL vs Logn & -5.9109 & 0.0013 & -3.2245 & 238 & \\
              &      & Logn vs Exp & 520.2079 & 0.0000 & 5.4694 & 238 & \\
United States & 2019 & PL vs Exp & 279.8822 & 0.0000 & 6.9651 & 341 & Power law or Lognormal \\
              &      & PL vs Logn & -1.0312 & 0.2974 & -1.0420 & 341 & \\
              &      & Logn vs Exp & 280.9134 & 0.0000 & 7.1160 & 341 & \\
\bottomrule
\end{tabular}
}
\end{table}

\newpage
\paragraph*{S4 Figure.}
Comparison of fits of play count distributions (PL vs Exp vs Logn.).

\begin{figure}[!h]
\centering
\includegraphics[width=1.0\linewidth]{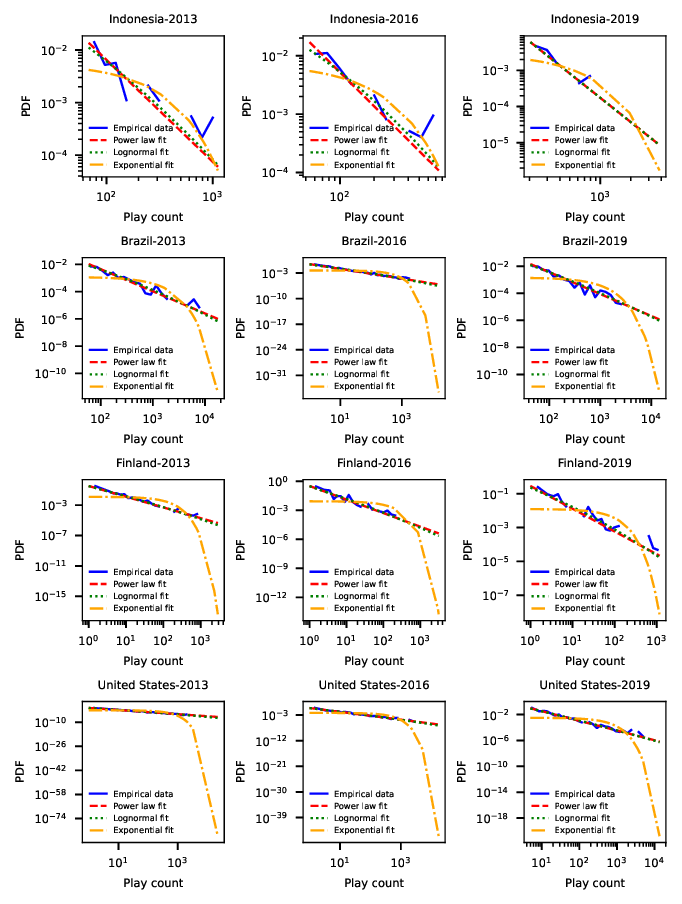}
\label{fig_s4}
\end{figure}

\end{document}